\documentclass[aps,twocolumn,prd,showpacs,nofootinbib]{revtex4}
\usepackage{amsmath}
\usepackage{graphicx}
\usepackage{dcolumn}
\usepackage{bm}
\usepackage{amssymb}
\usepackage{latexsym}

\def\be{\begin{equation}}
\def\ee{\end{equation}}
\def\ba{\begin{eqnarray}}
\def\ea{\end{eqnarray}}

\begin{document}

\title{The Primordial Perturbation Spectrum from Various Expanding and Contracting
Phases
%Before the``Bounce"
}

\author{Yun-Song Piao$^{a,b}$}
\email{yspiao@itp.ac.cn}
\author{Yuan-Zhong Zhang$^{c,d}$}
\affiliation{${}^a$Institute of High Energy Physics, Chinese
Academy of Science, P.O. Box 918-4, Beijing 100039, P. R. China}
\affiliation{${}^b$Interdisciplinary Center of Theoretical
Studies, Chinese Academy of Sciences, P.O. Box 2735, Beijing
100080, China} \affiliation{${}^c$CCAST (World Lab.), P.O. Box
8730, Beijing 100080} \affiliation{${}^d$Institute of Theoretical
Physics, Chinese Academy of Sciences, P.O. Box 2735, Beijing
100080, China }

%\date{Mar. 20$^\mathrm{th}$, 2004}

\begin{abstract}

In this paper, focusing on the case of single scalar field, we
discuss various expanding and contracting phases generating
primordial perturbations, and study the relation between the
primordial perturbation spectrum from these phases and the
parameter $w$ of state equation in details. Furthermore, we offer
an interesting classification for the primordial perturbation
spectrum from various phases,
% In this paper we describe the
%evolution of various expanding and contracting phases before the
%"bounce" in Einstein frame. Focusing on the case of simple scalar
%field, in term of the parameter $w$ of state equation, four
%different regions denoting different expanding and contracting
%phases are pointed out. We discuss possible scenarios generating
%the nearly scale-invariant spectrum and show that the inflation
%and other alternative scenario recently proposed can be placed in
%different positions of these regions. Our work offers an
%interesting classification for the primordial perturbation
%spectrum from various phases before the "bounce",
which may have
important implications for building an early universe scenario
embedded in possible high energy theories.
\end{abstract}

\maketitle

%During the last decade, observational cosmology has made enormous
%progresses. Due to the important role of primordial perturbation
%on formation of cosmological structure, it is required and urgent
%to understand the origin and nature of primordial perturbation.

Due to the central role of primordial perturbations on the
formation of cosmological structure, it is important probe their
possible nature and origin. The basic idea of inflation is simple
and elegant \cite{GL}, for a recent review, see Ref. \cite{LIN}.
%in which the scale factor expand very rapidly and the
%Hubble horizon is near constant, thus the causal horizon becomes
%much larger than the Hubble horizon and the initial perturbation
%in the horizon get amplified and exit the horizon during
%inflation, and freeze in as classical fluctuation in the energy
%density.
%However, the standard scenario of inflation dose not
%emerge in a direct way from any high energy physics theory. This
%makes it very flexible, which may be one of the main reasons why
%the basic scenario has survived for so long. For a given model
%which do not work, one can always make it work by slightly
%changing the potential or couplings of the scalar field, which
%straightly lead to a lot of different inflation models presented
%in various literatures. This flexibility may be regarded as a good
%point but in some sense a drawback.
A lot of observations, specifically the recent WMAP results
\cite{BHK}
%such as the flatness of space, the nearly
%scale-invariance, adiabaticity and Gaussian of the density
%perturbations
imply that the inflation is very consistent early cosmological
scenario.
%but this do not means that the inflation is uniquely
%consistent scenario with the observations.
However there remains some alternatives. The ekpyrotic/cyclic
scenario \cite{KOS, GKS, KST, TTS} is motivated by the string/M
theory.
%in which the
%visible universe is a boundary brane in five dimensional bulk
%space-time and the collision between two boundary branes leads to
%the reheating in visible universe corresponding to the Big Bang of
%standard cosmology.
The relevant dynamics with primordial perturbations can be
described by a 4D effective theory in which the separation of the
branes in the extra dimensions is modeled as a scalar field, and
dependent on the matching conditions \cite{TTS, D, CDC}, its
primordial perturbations spectrum may be nearly scale-invariant,
for some criticisms see Ref. \cite{BF, H, TBF}.
%In this scenario, the perturbation
%leaving the horizon during the contracting phase reenters the
%horizon after the bounce to an expanding phase corresponding to
%our observational cosmology. If the proper matching conditions
%during the bounce is considered \cite{TTS, D, TT}, see also
%\cite{PP, CDC}, the nearly scale-invariant spectrum could be
%obtained.
Furthermore, for a contracting phase like Pre Big Bang scenario
\cite{GV, V}, there is another case to seed a scale-invariant
spectrum \cite{FB} in which the pressureless matter is used, but
the corresponding scale solution is not an attractor \cite{GKS,
EWST}. For an expanding phase, in addition the usual inflation
scenario which gives scale-invariant spectrum,
%recently, we have
%proposed
a slowly expanding scenario \cite{PZ} with the phantom matter
($w<-1$) may be also feasible.
%\footnote{The matter with the state parameter $\omega
%\equiv {p\over \rho}<-1$, dubbed "phantom matter", has received
%increased attention recently. }, in which the 4D scale factor
%expands very slowly and is nearly constant, the Hubble parameter
%increase gradually. We find that when this slowly expanding phase
%is matched to another expanding phase with usual radiation and
%matter by some mechanism, a nearly scale-invariant spectrum may be
%generated.

All these scenarios rely on the parameter $w\equiv {p\over \rho}$
of state equation having a specific qualitative behavior
throughout the period when the perturbations are generated. For
the inflation scenario, the condition on $w$ is $w\simeq -1 $, and
power-law inflation with $a\sim t^n$ firstly studied in Ref. \cite
{AW}, it is $-1<w <-{1\over 3}$, and for ekpyrotic/cyclic
scenario, $w\gg 1$, and for slowly expanding scenario, $w\ll -1$.
%Correspondingly, the Hubble
%parameter is nearly constant during inflation, and the 4D scale
%factor contracts very slowly and is nearly constant during
%ekpyrotic/cyclic, and the 4D scale factor expands very slowly and
%nearly constant during slowly expanding scenario.
In some sense, all these scenarios can give some results
satisfying the WMAP observations in their simplest realization.
%The primordial perturbations of power-law inflation, in which
%$a(t)\sim t^n~~ (n>1)$ and $-1<w <-{1\over 3}$, have been studied
%firstly in Ref. \cite{AW}.
In this paper, focusing on the case of single scalar field, we
discuss various expanding and contracting phases with constant
$w$, and study the relation between the primordial perturbation
spectrum from these phases and the parameter $w$ of state equation
in details. Furthermore, we offer an interesting classification
for the primordial perturbation spectrum from various phases.
%and discuss possible scenarios
%generating the nearly scale-invariant perturbation spectrum.
%which is dependent
%on the matching condition crossing the "bounce".

\begingroup
%\squeezetable
\begin{table*}
\begin{tabular}{||p{3.2cm}|p{2.5cm}|p{2.8cm}|p{2.5cm}|p{2.5cm}||}\hline \hline
Region & I & II & III & IV \\
\hline the evolution of the scale factor & $a(t)\sim (-t)^n$ & $a(t)\sim t^n$ & $a(t)\sim (-t)^n$ & $a(t)\sim (-t)^n$ \\
\hline  & $n<0$ & $n>1$ & ${1\over 3}<n<1$ & $0<n<{1\over 3}$ \\
\hline & expansion & expansion & contraction & contraction \\
\hline the parameter of state equation & $w <-1$ & $-1<w <-{1\over 3}$ & $-{1\over 3} < w <1$ & $w >1$ \\
\hline kinetic energy term & reverse & standard & standard & standard \\
\hline
potential energy term & standard & standard & standard & reverse \\
\hline \hline
\end{tabular}
\caption{The natures of various expanding and contracting phases
generating the primordial perturbation spectrum are implemented in
single scalar field action. }
\end{table*}
\endgroup

In general the evolution of cosmological scale factor before the
``bounce" \footnote{Here the ``bounce" means the exit from
pre-bounce expanding/contracting phase to observational cosmology.
For inflation scenario it is the usual reheating \cite{KLS, FKL}.
} in Einstein frame can be written as \be a(t)\sim t^n
\label{at}\ee in which $t\rightarrow +\infty$, or \be a(t)\sim
(-t)^n \label{ant}\ee in which $t\rightarrow 0_-$, $n$ is constant
and positive or negative. For $n>0$, the (\ref{at}) corresponds
the expanding phase and the (\ref{ant}) corresponds the
contracting phase, and for $n<0$ the case is in reverse. The
Fridmann equations are \be h^2\equiv ({{\dot a}\over a})^2={\rho
\over 3}\label{h} \ee \be {\dot h}= - {\rho +p\over 2}\label{doth}
\ee where $8\pi G =1$ is set. Combing (\ref{at}), (\ref{ant}),
(\ref{h}) and (\ref{doth}), the power-law index of scale factor
\be n={2\over 3(1+w)}\label{w}\ee is given. For the case that the
speed of sound $c_s^2$ is constant the causally primordial
perturbations can be generated in such a phase, {\it i.e.} exits
the horizon during the evolution of this phase and re-enters the
horizon after the ``bounce" to an expanding phase corresponding to
our observational cosmology, which requires that $ah$ increases
with time, thus $n>1$ for (\ref{at}) and $n<1$ for (\ref{ant})
must be satisfied.

These phases can be implemented in the single scalar field action
as follows \be {\cal L}=-\epsilon\, {(\partial_\mu \varphi)^2\over
2} - V(\varphi) \label{action}\ee where $\epsilon$ is the sign of
the kinetic energy term, takes $1$ for normal scalar field and
$-1$ for phantom field. In this case $c^2_s =1 $. If taking the
field $\varphi$ spatially homogeneous but time-dependent,
%the energy density and the
%pressure are \be \rho = \mp {{\dot \varphi}^2 \over 2}+ V(\varphi)
%~~~~~~p= \mp {{\dot \varphi}^2 \over 2}-V(\varphi)\label{rhop}\ee
%and for phantom field
%$\rho = -{1\over
%2}{\dot \varphi}^2 + V(\varphi)$ and $p= -{1\over 2}{\dot
%\varphi}^2-V(\varphi)$.
%From (\ref{doth}) and (\ref{rhop}),
\be {\dot h}=-{n\over (-t)^2}=-\epsilon\, {{\dot \varphi}^2\over
2} \label{dotp}\ee can be given, which determines the nature of
scalar field, {\it i.e.} normal scalar field for $n>0$ or phantom
field for $n<0$. From (\ref{h}) and (\ref{dotp}), the effective
potential of the scalar field can be obtained and its pre-factor
is $n(3n-1)$, which determines the positive and negative of the
effective potential. Some details can be seen in Ref. \cite{PZ}.

\begin{figure}[t]
\begin{center}
\includegraphics[width=8cm]{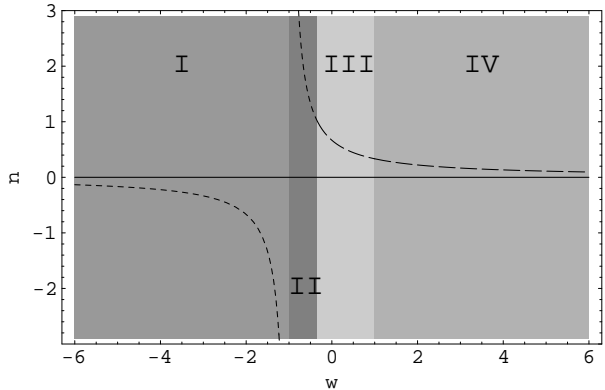}
\caption{The dot dashing line denotes the expanding phase and the
long dashing line denotes the contracting phase. From left to
right, the different shadows correspond to Region I II III and IV
in table respectively.} \label{fig1}
\end{center}
\end{figure}

Therefore, four different regions can be plotted according to the
relation between $n$ and $w$, see Fig. 1. Both regions I and II
are expanding phases. In region I, $a(t)\sim (-t)^n$ and $n<0$.
Since in this phase $w<-1$, the phantom field which has reverse
sign in kinetic energy term must be introduced. For $w\ll -1$, the
expansion is very slow but for $w\sim -1$, is very rapid, which
may be regarded as the phantom inflation \cite{PZH}. In region II,
$a(t)\sim t^n$ and $n>1$, which corresponds an accelerated
expanding phase. Such phase is generally described as power-law
inflation \cite{AW} and its extreme case $w\sim -1$ is usual de
Sitter inflationary phase. Different from region I and II, region
III and IV are contracting phases. In region IV, $a(t)\sim (-t)^n$
and $0<n<{1\over 3}$, compared with region III, which corresponds
a more slowly contracting phase. Since the pre-factor of the
potential is $n(3n-1)$, we see that in this region the negative
potential is required. The relevant details of these regions are
summarized in Table I. For the expanding phase, the scale solution
is a stable attractor only if $w<1$, which is compatible with
region I and II in which the scalar potentials are positive, and
for the contracting phase, the scale solution is a stable
attractor only if $w>1$, which is compatible with region IV, in
which the scalar potential is negative \cite{GKS, EWST}, but not
with region III.

In the following we discuss the primordial perturbation spectrums
from these regions before the ``bounce". Let us pay attention to
the scalar metric fluctuations. In longitudinal gauge and in
absence of anisotropic stresses, the scalar metric perturbation
can be written as \be ds^2 = a^2(\eta) (-(1+2\Phi)d\eta^2
+(1-2\Phi) \delta_{ij}dx^i dx^j )\ee where $\eta$ is conformal
time $d\eta \equiv{dt\over a}$, thus \be -\eta\sim (\pm t)^{-n+1}
\label{eta} \ee \be a(\eta) \sim (-\eta)^{{n\over 1-n}}\equiv
(-\eta)^q \label{aeta}\ee \be {\cal H}\equiv {a^\prime \over
a}\label{calh}\ee and $\Phi$ is the Bardeen potential.
% is \be
%\Phi^{\prime\prime} +2({\cal H}-{\varphi^{\prime\prime}\over
%\varphi^\prime})\Phi^\prime +2({\cal H}^\prime -{\cal
%H}{\varphi^{\prime\prime}\over \varphi^\prime})
%-\bigtriangledown^2\Phi=0 \ee
Defining the canonical variable \be u= {a\over \sqrt{2|{\cal
H}^2-{\cal H}^\prime |}}\Phi ={a\over \varphi^\prime}\Phi \ee For
adiabatic perturbation of single scalar field, the $u_k$ equation,
{\it i.e.} the linear perturbation equation of $\Phi$, can be
written as \be u_k^{\prime\prime}+\left(k^2 - {q(q+1)\over
\eta^2}\right)u_k =0 \label{uk}\ee  For all interesting modes $k$,
we can solve (\ref{uk}) analytically and obtain \be u_k =
\sqrt{-k\eta}\left( B_1(k)J_\mu
(-k\eta)+B_2(k)J_{-\mu}(-k\eta)\right)\label{uks}\ee where
$\mu=|q+{1\over 2}|$ and $J_\mu$ is the first kind of the Bessel
function with order $\mu$ and the function $B_i(k)$ can be
determined by specifying the initial conditions. In the regime
$k^2\eta^2 \gg q(q+1)$, in which the mode $u_k$ is very deep in
the horizon, the equation (\ref{uk}) reduced to the equation for a
simple harmonic oscillator, and $ u_k \sim {e^{-ik\eta} \over
(2k)^{3/2}}$ is stable. In the regime $k^2\eta^2 \ll q(q+1)$, in
which the mode $u_k$ is far out the horizon, the mode is unstable
and grows. In long-wave limit, $\Phi_k$ can be given and expanded
to the leading term of $k$ \be k^{3\over 2} \Phi \sim k^{{1\over
2}-\mu}\ee Thus the spectrum index from $\Phi$ is $n_{\Phi} = 1-2q
$ for $q > -{1\over 2}$ and $n_{\Phi} = 3+2q$ for $q<-{1\over 2}$.

The curvature perturbation on uniform comoving hypersurfaces \be
\zeta ={{\cal H}\Phi^\prime +{\cal H}^2 \Phi\over {\cal H}^2
-{\cal H}^\prime}+\Phi \label{zeta}\ee is a constant on scales
larger than Hubble horizon in the absence of entropy fluctuations,
as can be seen from its equation of motion \be \zeta^\prime =
{{\cal H}\over {\cal H}^2-{\cal H}^\prime } k^2 \Phi
\label{dzeta}\ee and is used to infer the spectrum of $\Phi$ at
the time when the perturbations re-enter the Hubble horizon in
inflationary cosmology. Defining the Mukhanov-Sasaki variable
\cite{M, MFB}
%\footnote{
%We shall see that the spectrum of $\Phi$
%and $\zeta$ obtained from $u$ and $v$ is not equivalent, which
%case is correct can not be decided on the level of the equations
%of motion. But considering
%$v$the action for scalar metric
%fluctuation,  may be regarded as a proper quantity that should
%be quantized \cite{M, MFB}. }
\be v= {a\sqrt{2|{\cal H}^2 -{\cal H}^\prime |}\over {\cal H}}
\zeta ={a\varphi^\prime\over {\cal H}} \zeta \ee The $v_k$
equation can be written as \be v_k^{\prime\prime}+\left(k^2
-{q(q-1)\over \eta^2}\right)v_k =0 \ee Similarly we can obtain \be
v_k = \sqrt{-k\eta}\left( C_1(k)J_\nu
(-k\eta)+C_2(k)J_{-\nu}(-k\eta)\right)\label{uks}\ee where
$\nu=|q-{1\over 2}|$ and $J_\nu$ is the first kind of the Bessel
function with order $\nu$ and the function $C_i(k)$ can be
determined by specifying the initial conditions. The initial
condition is $v_k \sim {e^{-ik\eta} \over (2k)^{1\over 2}}$. In
long-wave limit, $\zeta_k$ can be given and expanded to the
leading term of $k$ \be k^{3\over 2} \zeta \sim k^{{3\over 2}-\nu}
\ee Thus the spectrum index from $\zeta$ is $n_{\zeta} = 5-2q $
for $q
> {1\over 2} $ and $n_{\zeta} = 3+2q $ for $ q<{1\over
2}$.

\begin{figure}[t]
\begin{center}
\includegraphics[width=8cm]{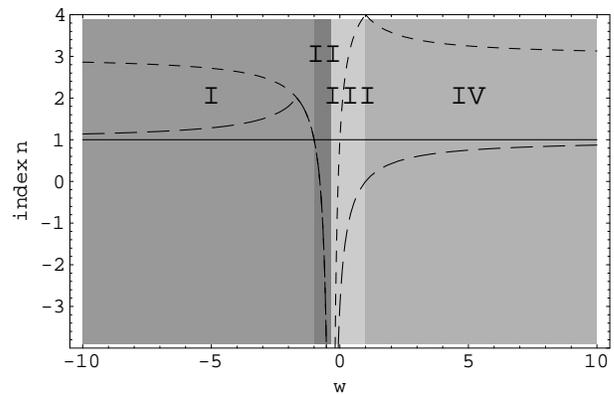}
\caption{The short dashing line denotes $n_{\zeta}$, the long
dashing line denotes $n_{\Phi}$. From left to right, the different
shadows correspond to Region I II III and IV in table
respectively. } \label{fig2}
\end{center}
\end{figure}

\begin{figure}[t]
\begin{center}
\includegraphics[width=8cm]{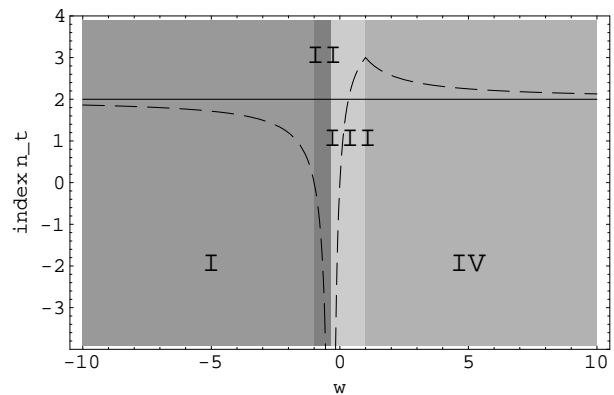}
\caption{The dashing line denotes $n_{t}$. From left to right, the
different shadows correspond to Region I II III and IV in table
respectively. } \label{fig3}
\end{center}
\end{figure}

Therefore, in general, for various expanding and contracting
phases the values of $n_{\Phi}$ and $n_{\zeta}$ are different.
%Fig. 1 is plotted.
%The red line denotes $n_{\Phi}$ and the green line denotes
%$n_{\zeta}$, when $n<{1\over 2}$, both lines overlap. In Fig. 1,
%different shadows means different regions, which, from left to
%right, correspond to Region I II III and IV in table respectively.
From (\ref{aeta}) and (\ref{calh}), \be {{\cal H}^\prime \over
{\cal H}^2}= -{1+3w \over 2} \ee can be obtained, thus \be
q={2\over 1+3w} \label{n} \ee In term of relations between
$n_{\Phi}$ or $n_{\zeta}$ and $q$, the Fig. 2 reflecting
$n_{\Phi}$ or $n_{\zeta}$ varying with $w$ is plotted. $n_t$
\footnote{The tensor perturbed metric can be written as \be ds^2 =
a(\eta)(- d\eta^2+(\delta_{ij}+h_{ij})dx^i dx^j)\ee where $h_{ij}$
can be expanded in term of the two basic traceless and symmetric
polarization tensors $e_{ij}^+$ and $e_{ij}^\times$ as
$h_{ij}=h_{+}e_{ij}^+ + h_{\times} e_{ij}^\times$. The
gauge-invariant tensor amplitude $\mu_k ={a\over \sqrt{2}}h_k$
satisfies the equation \cite{SAW} \be
\mu_k^{\prime\prime}+(k^2-{q(q-1)\over \eta^2})\mu_k =0\ee Thus
the tensor spectrum index is $n_t =4-2q$ for $q>{1\over 2}$ and
$n_t =2+2q$ for $q<{1\over 2}$. }is the spectrum index of the
tensor fluctuations and plotted in Fig. 3.
%From left to right,
%the corresponding shadow regions denote the region I, II, III and
%IV.
We see that only for $-{5\over 3}<w<-{1\over 3}$, {\it i.e.}
$q<-{1\over 2}$, the spectrum index of $\zeta$ is the same as that
of $\Phi$, and when $w\simeq -1$, nearly scale-invariant spectrum
can be obtained, which corresponds to the case in inflationary
cosmology, but for other value of $w$, the spectrum index of
$\zeta$ is different from that of $\Phi$, the $\Phi$ spectrum is
nearly scale-invariant for $w \ll -1$ and $\gg 1$, while the
$\zeta$ spectrum is nearly scale-invariant for $w \simeq 0$. Which
of the spectrum of $\Phi$ and $\zeta$ can be inherited in
late-time observational cosmology dependent on the matching
conditions through the ``bounce" \cite{D}. We shall briefly
comment it in the following.
%If the $\Phi/\zeta$ mixing regularly crosses the
%"bounce", the nearly scale-invariant spectrum requires $w\simeq
%-1, \ll -1$ and $\gg 1$, for $w\simeq 0$, although $\zeta$ is
%nearly scale-invariant, $\Phi$ acquires a strong red spectrum,
%thus the final mixing spectrum will be unacceptable.

The curvature perturbations re-enter the horizon during
radiation/matter-domination and create density fluctuations
$\delta \rho$, which results in the formation of large-scale
structure of universe. The density contrast $\delta ={\delta
\rho\over \rho}$ can deduced from Poisson equation \be \delta_k
={2\over 3} ({k\over {\cal H}})^2 \Phi_k \ee at horizon crossing
${\cal H}\simeq k $, $|\delta_k|^2 \simeq |\Phi_k|^2$ is given,
thus the Bardeen potential in late-time universe really determines
the spectrum. The solution for $\Phi$ on super Hubble scale is
approximately \cite{MFB} \be \Phi_{\pm}\simeq D_{\pm}
+S_{\pm}{{\cal H}\over a^2} \label{Phim}\ee From (\ref{zeta}) and
(\ref{dzeta}), we have \be \zeta_{\pm} = \alpha D_{\pm} + \beta
k^2 S_{\pm} f(\eta) \label{zetam}\ee where the subscript $+$ and
$-$ denote the phases after and before the ``bounce" respectively,
the coefficient $D_- \sim k^{-{1\over 2} +q}$ and $D_+$ denote the
amplitude of the constant mode, $S_-\sim k^{-{3\over 2} -q}$ and
$S_+$ are that of the growing or decaying mode dependent on
different phases, and $f(\eta) =\int ({{\cal H}\over
a\varphi^\prime})^2 d\eta$, the coefficients $\alpha$ and $\beta$
are dependent only on the parameter $w$ of state equation. For
late-time expanding phase driven by radiation/matter, $S_+$ is
decaying mode, thus $D_+$ is dominated mode for our observational
cosmology. If taking the constant energy hyper-surfaces, in which
$\Phi$ and $\zeta$ are continuous across the bounce, {\it i.e.}
the Deruelle-Mukhanov \cite{DM} (Hwang-Vishniac \cite{HV})matching
conditions is satisfied, like that in the transition from
inflationary phase to radiation dominated phase, $D_+ \sim D_-$
can be obtained to leading order of $k$ \cite{BF}, thus regardless
which of $D_-$ and $S_-$ is dominated mode during pre-bounce
phase, $D_+$ mode will only inherit the spectrum of $D_-$ mode. In
this case we can obtain the same results from $\Phi$ and $\zeta$.
But if, for example, we take $\Phi$ and $\Phi^\prime$ continuous,
then $D_+ \sim D_- + S_- $, which means that $D_+$ mode should
take the spectrum of dominated mode which of $D_-$ and $S_-$. For
ekpyrotic/cyclic and slowly expanding scenario, the growing $S_-$
mode is the dominated mode during pre-bounce phase, which will be
inherited by $D_+$ mode. From (\ref{Phim}) and (\ref{zetam}), we
see that for $S$ mode, the spectrum of $\Phi$ is different from
that of $\zeta$ and the latter has a suppression from $k^2$, which
may imply that $\zeta$ is not a proper quantity for the
calculations of primordial perturbation spectrum in this case.
More recently, as is pointed to in Ref.\cite{CDC}, see also Ref.
\cite{PP, MP}, the resulting spectral index in late
radiation-dominated universe depends on how $\Phi$ and $\zeta$
passing through the ``bounce",
%{\it i.e.} the nearly
%scale-invariant spectrum can be obtained
%when $w\simeq -1$, $\ll -1$ and $\gg 1$ for
%regular $\Phi$ or $\Phi/\zeta$ and $w\simeq -1$ and $0$ for regular $\zeta$,
which is determined by the details of ``bouncing" physics.
%and may
%lead to an uncertainty for the perturbation spectrum.
%The effect of this
%uncertainty on the
%spectrum of pre-bounce is debated. However, a more generic
%classification including the effect of "bounce" may be more
%interesting, which is worth studying further. We will go back to
%these issues in the future.
%In this paper, we neglect the
%possible effects of "bounce" on the pre-bounce metric
%fluctuations. In Ref. \cite{MP}, this effect has been considered,
%in which the curvature term is introduced to smooth the bounce.

%We have assumed that the transition will occur when the scalar
%field decays into the usual radiation by some coupling, or through
%the singular "Big Rip" or "Big Bang" into the radiation-dominated
%phase by some other mechanism from the string/M theory or other
%high energy/dimension theory, which may be regarded as a "bounce"
%to our late-time observational cosmology. For the expanding phase
%I and II, the decay of the field driving expansion may be a
%feasible exit mechanism, which is similar to reheating after
%inflation to the extent. For the contracting phase III and IV, in
%addition the high order corrections and through the singularity,
%introducing the curvature term \cite{GT, MP} and extra negative
%energy field \cite{PP, F} also can smooth the bounce.

In summary, we describe the evolution of various pre-bounce phases
with constant $w$ and also construct the corresponding action of
the single scalar field. We separate four different regions which
correspond different expanding and contracting phases in term of
parameter $w$ of state equation, and show that the inflation and
other alternative scenarios recently proposed can be placed in
different positions of these regions, and for all possible $w$,
only when $w\simeq -1$ or $\ll -1$ or $\gg 1$ may the nearly
scale-invariant scalar spectrum be obtained, which corresponds the
inflation, ekpyrotic/cyclic and slowly expanding scenario
respectively.
%For $w\sim 0$, although $\zeta$ is nearly
%scale-invariant, $\Phi$ is strong red, which may lead an
%unacceptable large perturbation.
The degeneration of nearly scale-invariant scalar fluctuations
spectrum may be remove by the tensor fluctuations spectrum, see
Fig. 3, which, for $w\simeq -1$, is nearly scale-invariant and for
$w\ll -1$ or $\gg 1$, is strong blue. Since different scenarios
make different predictions for the spectrum of scalar and tensor
perturbations, it is possible that the observations of the cosmic
microwave background can distinguish among different scenarios.
For other value of $w$, the nearly scale-invariant scalar spectrum
can not be generated by the fluctuations of background field, thus
other feasible mechanisms may be required \cite{LW, BGGV, DGZ}.
Furthermore, the matching between various phases \cite{PFZ, PTZ}
may give a reasonable explain for a possible loss of power
recently observated by WMAP.
%In this work, we only focus on the case of simple field. The
%generalization to multi-field remains. We also neglect the detail
%of "bounce", since its effect
Our work offers an interesting classification for the primordial
perturbation spectrum from various phases before the ``bounce",
which may have important implications for building an early
universe scenario embedded in possible high energy theories.

\textbf{Acknowledgments}
%We would like to thank the innominate
%referees for kind comments and suggestions.
We thank Mingzhe Li, Xinmin Zhang for useful discussions. We also
thank Robert Brandenberger for reminding on the matching
conditions in Ref. \cite{BF}. When this manuscript was being
submitted to the relevant journal, Boyle {\it et.al.}'s paper
appeared \cite{BST}, in which the scalar pertrubations ($\Phi$ and
$\zeta$) and tensor perturbations as a function of $w$ were also
obtained, furthermore, they pointed out the interesting
relationship between two different models sharing the same scalar
perturbations. We thank its authors for kind correspondence and
Paul J. Steinhardt for comments on our manuscript. This work is
supported in part by K.C.Wang Postdoc Foundation and also in part
by the National Basic Research Program of China under Grant No.
2003CB716300.

\end{document}